\documentclass[pra,twocolumn,showpacs,preprintnumbers,amsmath,amssymb,
]{revtex4}
\usepackage{graphicx}
\usepackage{color}

\begin{document}
\title{\bf Optimal minimum-cost quantum measurements for imperfect detection}
\author{Erika Andersson}
\address{SUPA, Department of Physics, Heriot-Watt University, Edinburgh EH14 4AS, United Kingdom}

\date{\today}
\begin{abstract}
Knowledge of optimal quantum measurements is important for a wide range of situations, including quantum communication and quantum metrology.
Quantum measurements are usually optimised with an ideal experimental realisation in mind. Real devices and detectors are, however, imperfect. This has to be taken into account when optimising quantum measurements. 
In this paper, we derive the optimal minimum-cost and minimum-error measurements for a general model of imperfect detection.
\end{abstract}
\maketitle

\section{Introduction}
Quantum measurements may be optimized with respect to a range of criteria. For example, when distinguishing between a set of quantum states $\rho_j$, occurring with prior probabilities $p_j$, we may want to minimise the average error in the result, or, somewhat more generally, the average cost~\cite{helstrom, rehacek, steve}. 
Other possibilities are to maximise the information contained in the result, or to ask that measurement results should be unambiguous~\cite{rehacek}. 
The optimal measurement is often a generalized quantum measurement. These go beyond standard projective quantum measurements, and are  referred to as probability operator measures (POMs) or positive operator-valued measures (POVMs)~\cite{helstrom, rehacek, steve}. How to optimally perform a quantum measurement is relevant for a wide range of applications, from quantum communication including quantum key distribution~\cite{enk}, to parameter estimation and quantum metrology, such as for optimal quantum estimation of the Unruh-Hawking effect~\cite{aspachs}.

Derivations of optimal quantum measurement strategies have, however, aimed at finding the optimal {\it ideal} quantum measurement, that is, the best that can be done provided that an actual experimental realisation is ideal. An advantage is that the theoretical derivation of the optimal ideal measurement can be separated from working out how to do the experimental realization. The same theoretically optimal measurement may be realised in very different ways, depending on the character of the quantum system to be measured.
Generalized quantum measurements have been realized on photon polarisation, see e.g.~\cite{huttner, clarkeunamb, clarkesymm, mizuno, superadd, source, steinberg, mosley}, and very recently on NV-centres in diamond~\cite{stuttgart}. They could also be realized on ions or atoms with existing experimental tools~\cite{atompom, atompom2, cavity}. 

Real experiments are of course not ideal. One might ask whether imperfections in device components changes the strategy that we should aim to implement. This question has largely been overlooked. It turns out that the measurement we should aim to implement does indeed in general depend on the particular properties of the experimental devices used. An indication that this might be the case is given by unambiguous comparison of quantum states, where for detectors with less than unit efficiency, it is sometimes advantageous to use an amplifier in front of the detector~\cite{hamilton}. This is not self-evident, since an amplifier will add noise, which could degrade the performance of a measurement. 

In this paper, we  will derive the optimal minimum-cost measurement for the case when the final detection process is not perfect. It turns out that when distinguishing between two quantum states, the optimal measurement strategy we should aim for remains the same as in the ideal case, although the cost changes. For three or more states, the optimal measurement strategy in general changes, and we give an example of this. We finish with a discussion.

\section{generalized quantum measurements}
Generalized quantum measurements (POMs) can be realised in terms of a projective measurement in an extended Hilbert space~\cite{helstrom,steve}. This is used in both existing and suggested experimental realisations. 
The Hilbert space of the quantum system $\rho_S$ to be measured can either be extended through a direct sum, {\it e.g.} by using extra atomic levels or adding more optical paths, or through a tensor product by coupling $\rho_S$ to an auxiliary quantum system $\rho_A$. 
Broadly speaking, the number of dimensions in the total extended Hilbert space corresponds to the number of outcomes of the measurement. Less well known is that by realising the measurement sequentially, it is also possible to limit the total number of dimensions needed at any one time to $d+1$, where $d$ is the dimension of $\rho_S$~\cite{chinese}. We can also realise the measurement by using at most $2d$ dimensions, by coupling $\rho_S$ to an auxiliary qubit, measuring the qubit, and repeating~\cite{treepom}. This realisation is the most efficient in the sense that fewer operations are needed on average.

Any ideal experimental realisation of a generalized quantum measurement will thus employ a final projective measurement in some basis on some quantum system.  The final projective measurement is preceded by a unitary transform in the extended Hilbert space.  
For example, for a measurement on an atom or ion, we may couple the levels of the atom or ion to some additional atomic levels e.g. using laser pulses or passage through a cavity~\cite{atompom,cavity}, followed by a final measurement of which energy level the atom or ion occupies. Due to experimental constraints, the final measurement is restricted to be a projection in the energy eigenbasis. If making measurements on a photon, the final measurement might be a detection of which path the photon exits from, and with which polarisation, in a suitable polarisation basis~\cite{clarkeunamb, clarkesymm, mizuno, superadd,source, steinberg, mosley}. This final measurement is preceded by an optical network which includes wave plates and beam splitters, effecting a unitary transform in the extended space.

Formally, any generalized measurement strategy is described by a set of measurement operators $\Pi_i$, acting in the space of $\rho_S$, the system to be measured. If the system is prepared in a state $\rho_j$, then the probability to obtain outcome $i$ is given by $p(i|j)={\rm Tr}(\Pi_i\rho_j)$. The fact that the sum of the probabilities for all outcomes is equal to 1 corresponds to the condition $\sum_i\Pi_i={\bf 1}$, and the fact that probabilities for all outcomes are nonnegative corresponds to $\Pi_i \ge 0$ (and consequently $\Pi_i$ have to be Hermitian). Note that we may have more outcomes than we have dimensions in the space of the system to be measured, and that $\Pi_i$ need not be orthonormal. Each measurement operator $\Pi_i$ is obtained by taking the corresponding projective measurement operators in the extended Hilbert space and projecting them onto the subspace of $\rho_S$.

\section{Imperfect detection}
Sources of error in the experimental realisation can now be divided in two categories, errors in the unitary transform preceding the final measurement, and errors in the final projective measurement. In this paper, we will consider the latter type of errors. Errors in the unitary transform are of course also likely to occur in any realization, and this will be the subject of future work.

In particular, suppose that when a perfect projective measurement would have given result $j$, then our measurement will instead give result $i$ with probability $q(i|j)$, where $\sum_i q(i|j)=1$; $q(i|j)$ are the elements of a stochastic matrix. This is a very generally applicable description of errors in the final measurement. We may be able to vary the unitary transform that precedes the final measurement, but the final measurement and the $q(i|j)$ are often fixed by the nature of the chosen detection process, including the efficiency of the detectors, and what basis states we must project on.
Any non-ideal measurement we are able to implement will then be described by mixed measurement operators
\begin{equation}
\label{eq:imppom}
\widetilde{\Pi}_i = \sum_j q(i|j)\Pi_j,~~\sum_i q(i|j)=1,
\end{equation}
where $\Pi_j$ are the measurement operators for an unconstrained generalized measurement strategy, acting in the space of $\rho_S$. 

For example, when trying to implement a photon number measurement, the measurement that we are actually able to implement might be described by the measurement operators
\begin{eqnarray}
\Pi_{\it no~click} &=& \sum_{n=0}^\infty q(0|n)|n\rangle\langle n| \nonumber\\
\Pi_{\it click} &=& \sum_{n=0}^\infty q(1|n) |n\rangle\langle n|,
\end{eqnarray}
where $|n\rangle$ is a photon number state. This measurement thus has only two outcomes, and $q(0|n)$ and $q(1|n)$ are the probabilities to obtain ``no click" and a ``click" respectively, when there where $n$ photons present. A similar description results for photon-number-resolving detectors, and also when a number of photodetectors are used to detect what path or with what polarisation a photon exits. Yet another example is the detection of the state of a Rydberg atom by field ionization. By detecting for what field strength the atom is ionized one can infer what energy level it most likely would have been in~\cite{fieldionization,cavity}. The distribution functions for when the electron is released may however overlap for different energy levels. This also results in a measurement of the type in Eq. \eqref{eq:imppom}.

There is a minor technical point which should be mentioned in order to further justify the  generality of our error model. If the measurement operators $\Pi_j$ are proportional to pure state projectors, then the final measurement in the extended space can be chosen as a projection in a complete basis, with each outcome corresponding to one pure basis state $|j\rangle$. Our model for errors in the final measurement is then natural. The measurement operators $\Pi_j$ may however also be mixed. This is of course possible to realise using a final measurement comprising projectors onto more than one orthonormal state. However, the index $j$ in $q(i|j)$ most naturally refers to pure-state projectors $|j\rangle\langle j|$ in the final measurement, rather than to projectors onto more than one orthonormal state.

Nevertheless, also for mixed $\Pi_j$, we may arrange for the corresponding final measurement operator to be a pure state projector in an extended space. This means that the description using $q(i|j)$ to describe misidentification proabilities in the final measurement is again natural. For example, suppose that the measurement in the extended Hilbert space is described by the projectors  $ |1\rangle_{EE}\langle 1|+|2\rangle_{EE}\langle 2|$ and $|3\rangle_{EE}\langle 3|$. We may then further couple the system to a qubit, initially in the state $|0\rangle_q$, using e.g. the unitary operation
\begin{eqnarray}
U&=&|1\rangle_{EE}\langle 1|\otimes|0\rangle_{qq}\langle 0| +|2\rangle_{EE}\langle 2|\otimes|0\rangle_{qq}\langle 0|\nonumber\\ 
&+&|1\rangle_{EE}\langle 1|\otimes|1\rangle_{qq}\langle 1| +|2\rangle_{EE}\langle 2|\otimes|1\rangle_{qq}\langle 1|\nonumber\\ 
&+&|3\rangle_{EE}\langle 3|\otimes|1\rangle_{qq}\langle 0|+|3\rangle_{EE}\langle 3|\otimes|0\rangle_{qq}\langle 1|.
\end{eqnarray}
The final measurement may then be realized as a projective measurement on the qubit, with $|0\rangle_{qq}\langle 0|$ corresponding to $ |1\rangle_{EE}\langle 1|+|2\rangle_{EE}\langle 2|$ and $|1\rangle_{qq} \langle 1|$ corresponding to $|3\rangle_{EE}\langle 3|$. This procedure is easily generalized so that any mixed measurement operators $\Pi_j$ will correspond to a pure state projector in some extended space. 
Related to this, one realizes that 
it may well be advantageous to arrange to use 
as few final measurement states as possible. Any extra unitary transforms are of course likely to introduce experimental errors. But in this initial work we want to optimize with respect to imperfections in the final projective measurement only.

\section{Minimum-cost measurements for imperfect detection}
We will now derive the optimal minimum-cost strategy when the final measurement we can realize is restricted, so that the measurement operators are given by Eq.  \eqref{eq:imppom}. The measurement we are actually realizing is described by the measurement operators $\widetilde{\Pi}_i$, whereas the measurement we are aiming to realize is described by the measurement operators $\Pi_j$.

To briefly review results related to optimal minimum-cost measurements \cite{helstrom}, suppose that quantum state $\rho_j$ occurs with prior probability $p_j$. We choose a measurement with measurement operators $\Pi_i$, and obtaining result $i$ when the state prepared was actually $\rho_j$ carries a cost of $C_{ij}$. 
The average cost will then be given by
\begin{equation}
\bar{C}={\rm Tr}\sum_{ij} C_{ij}p_j\Pi_i\rho_j
={\rm Tr}\sum_i 
W_i \Pi_i = {\rm Tr} \Gamma,
\end{equation}
where
\begin{equation}
W_i = \sum_j C_{ij} p_j \rho_j \text{~~and~~} \Gamma = \sum_i W_i \Pi_i =\sum_i\Pi_i W_i
\end{equation}
are called the risk operator corresponding to result $i$ and the Lagrange operator, respectively. $\Gamma$ takes care of the constraint $\sum_i\Pi_i={\bf 1}$, and is Hermitian. For a minimum-error measurement we may choose $C_{ij}=-\delta_{ij}$ and $W_i = -p_i \rho_i$.
The minimum-cost measurement operators satisfy the conditions
\begin{eqnarray} 
\label{eq:helstrom1}
(W_i-\Gamma)\Pi_i = \Pi_i (W_i-\Gamma) &=&0 ~~\forall i\\
W_i-\Gamma&\ge& 0 ~~\forall i.
\end{eqnarray}

Consider now a mixed measurement strategy with measurement operators given by Eq.  \eqref{eq:imppom}.  The average cost for this measurement strategy will be
\begin{eqnarray}
\label{eq:impcost}
\widetilde{C} &=& {\rm Tr} \sum_i W_i \widetilde{\Pi}_i = {\rm Tr} \sum_{ijk} C_{ik} p_k \rho_k q(i|j) \Pi_j\nonumber\\
&=&{\rm Tr} \sum_{jk} \widetilde{C}_{jk} p_k \rho_k \Pi_j = {\rm Tr} \sum_j \widetilde{W}_j \Pi_j,
\end{eqnarray}
where
\begin{equation}
\label{eq:cmod}
\widetilde{C}_{jk} = \sum_i C_{ik} q(i|j) ~~\text{and}~~ \widetilde{W}_j = \sum_k \widetilde{C}_{jk} p_k \rho_k.
\end{equation}
It immediately follows that the ideal strategy we should aim to perform, if the final measurement has the misidentification probabilities $q(i|j)$, is optimal for the modified costs $\widetilde{C}_{jk}$ and modified risk operators $\widetilde{W}_j$. It is also clear that if the states $\rho_j$ are orthogonal, corresponding to ``perfectly distinguishable" classical states, then the measurement in the presence of misidentification probabilities does not change; it remains a projective measurement on (the subspaces of) the different $\rho_j$. The fact that the measurement strategy changes is in this sense a {\it quantum} feature.

We can freely choose which final measurement states are assigned to which initial state $\rho_i$, and should choose so that the obtained cost is optimal. This can be done by checking what the optimal measurement is for each possible assignment and picking the best one; there are $m!$ assignments if there are $m$ basis states. Roughly speaking, the most probable initial states should be associated with those basis states which we can identify most accurately. Related to this, if the final detection process is defective enough, then for some assignments it may happen that when a final outcome $i$ is obtained, this is more likely to have occurred as a result of another initial state $\rho_j$ than the initial state $\rho_i$ itself. As a particularly simple example, consider distinguishing with minimum error between the orthogonal states $|0\rangle$ and $|1\rangle$, occurring with probabilities $p_0$ and $p_1$.
We make an imperfect projection in the $\{|0\rangle , |1\rangle \}$ basis, with 
$\widetilde\Pi_0=q(0|0)|0\rangle\langle 0|+ q(0|1)|1\rangle\langle 1|, \widetilde\Pi_1=q(1|0)|0\rangle\langle 0|+ q(1|1)|1 \rangle\langle 1|$. 
If for example $q(0|1) p_1 > q(0|0) p_0$, then the final result ``$0$" is more likely to have occurred because the state was $|1\rangle$ rather than $|0\rangle$, and we should guess ``$|1\rangle$" even if we obtain result ``0". 
We should therefore also check what the optimal cost is for different reassignments of final outcomes to other states.

Nevertheless, checking all possible different assignments of outcomes, in order to obtain the overall optimal detection strategy for imperfect detection, is straightforward if there is a finite number of outcomes. We will proceed to look at examples.

\subsection{Distinguishing between two non-orthogonal states}

Suppose that we want to distinguish between $\rho_0$ and $\rho_1$, occurring with prior probabilities $p_0$ and $p_1$, with minimum cost, with misidentification probabilities $q(0|0), q(0|1), q(1|0), q(1|1)$. The cost of obtaining result $i$ when the prepared state was $j$ is $C_{ij}$, for $i,j=0,1$. The ideal measurement strategy we should try to implement is optimal for the modified risk operators
\begin{eqnarray}
\widetilde{W}_0 &=&  \widetilde{C}_{00}p_0\rho_0 +  \widetilde{C}_{01} p_1\rho_1,\nonumber\\
\widetilde{W}_1 &=&  \widetilde{C}_{10}p_0\rho_0 + \widetilde{C}_{11}p_1 \rho_1.
\end{eqnarray}
The optimal measurement for distinguishing between two non-orthogonal states with minimum cost was given by Helstrom~\cite{helstrom}. It is a projection in the eigenbasis of the operator
$\widetilde{O}=\widetilde W_0-\widetilde W_1.$
Using the first equation in Eq. \eqref{eq:cmod} and the second equation in Eq. \eqref{eq:imppom}, we find that
\begin{equation}
\widetilde{O}=[p(1|1)+p(2|2)-1](W_0-W_1).
\end{equation}
This means that unless $p(1|1)+p(2|2)=1$, the measurement strategy we should aim to implement does not change when detection is imperfect. If $p(1|1)+p(2|2)>1$, then the optimal $\Pi_0$ is a projector onto the eigenstates of $\widetilde{O}$ with negative eigenvalues, and $\Pi_1$ is a projector onto the eigenstates with positive eigenvalues. If there are any zero eigenvalues, then the corresponding eigenstates may be assigned to either result without changing the average cost. 
If the results are sufficiently ``scrambled", more precisely, when $q(1|1)+q(2|2) < 1$, we should still perform the same measurement, but with $\Pi_0$ and $\Pi_1$ swapped. If $q(1|1)+q(2|2)=1$, then this implies $q(2|2)=q(2|1)$ and $q(1|1)=q(1|2)$, {\it i.e.} that the measurement results are completely random. There is then no point in making a measurement at all. We should just pick the $\rho_i$ which gives the least average cost based on the prior probabilities and costs $C_{ij}$.

The minimum cost, given in Eq. \eqref{eq:impcost}, does of course increase. These results also hold for the special case of distinguishing between $\rho_0$ and $\rho_1$ with minimum error. The strategy we should try to implement stays the same, but the error probability increases. That this holds even when the misidentification probabilities $q(i|j)$ are not symmetric is not entirely intuitive.

\subsection{Distinguishing between three symmetric states}

We will now see that when distinguishing between three pure quantum states, the measurement strategy we should aim for may change when the detection is imperfect. Consider the three equiprobable states 
\begin{equation}
|\psi_1\rangle=-|0\rangle, ~|\psi_2\rangle= \frac{1}{2}(|0\rangle +\sqrt{3})|1\rangle, ~|\psi_3\rangle= \frac{1}{2}(|0\rangle -\sqrt{3})|1\rangle.
\end{equation}
The ideal measurement that distinguishes between these states with minimum error has the measurement operators $\Pi_i = 2/3|\psi_i\rangle\langle\psi_i|=2/3\rho_i$ for $i=1,2,3$~\cite{helstrom, rehacek, clarkesymm}.

Suppose now that in the final detection, outcome 1 is sometimes misidentified as outcome 2 or 3  with equal probability $q$, but that otherwise the detection is perfect. That is, we have $q(1|1)=1-2q$ and $q(2|1)=q(3|1)=q$, with $0 \le q \le 1/2$. Also, $q(2|2)=q(3|3)=1$ and $q(1|2)=q(3|2)=q(1|3)=q(2|3)=0$. We find $\widetilde C_{11}=2q-1, \widetilde C_{12}=\widetilde C_{13}=-q, \widetilde C_{22}=\widetilde C_{33}=-1$, and all other $\widetilde C_{ij}=0$. Furthermore,
\begin{eqnarray}
\widetilde W_1&=&\frac{1}{3}\left[(2q-1)\rho_1-q(\rho_2+\rho_3)\right],~~\nonumber\\
\widetilde W_2&=&-\frac{1}{3}\rho_2, ~~\widetilde W_3 =-\frac{1}{3}\rho_3.
\end{eqnarray}
The optimal measurements for such mirror-symmetric situations are known~\cite{mirrorsym, chou}. For small $q$, the optimal measurement strategy has three measurement operators. When $q$ increases, it pays less and less try to identify $|\psi_1\rangle$,  and the trace $a$ of $\Pi_1=a\rho_1$ decreases, starting from 2/3. At the same time,  ${\rm Tr}~\Pi_2={\rm Tr}~\Pi_3$ increases. $\Pi_{2,3}$ are unnormalised projectors onto pure states that become closer and closer to $|\pm\rangle = 1/\sqrt{2}(|0\rangle \pm |1\rangle)$. When $q\ge q_c=(1+1/\sqrt{3})/2\approx 0.211$, $a=0$ and the optimal measurement has only two non-zero measurement operators. Then $\Pi_1=0$, and $\Pi_{2,3}$ are projectors onto the states $|\pm\rangle$. This remains optimal for all $q_c\le q\le 1/2$. Thus, for any non-zero value of $q$, the measurement we should aim for is different from the optimal minimum-error measurement for $q=0$.

The case we have considered, distinguishing between three or more symmetric states, is relevant for quantum key distribution (QKD), see e.g.~\cite{sych} and references therein. As argued above, our detection error model directly applies to photodetection. In a realisation of similar QKD protocols, it is therefore likely that optimal operation would require similar modifications of the measurements performed.

\section{Conclusions}

Knowledge of optimal quantum measurements for realistic experimental components is important in order to be able to select the best possible measurements for a given situation.
In this work, we derived the optimal minimum-cost measurement one should aim to implement for a general model of imperfect detection. When a perfect measurement would have given result $j$, then  the detection gives result $i$ with probability $q(i|j)$. This leads to a modification of the costs of different outcomes, and hence the measurement strategy we should aim to implement in general changes. In the special case of distinguishing between two quantum states, the measurement stays the same, and only the cost changes. For three states, we gave a simple example, relevant e.g. for quantum key distribution, where the optimal measurement strategy changes significantly if the detection is imperfect.

It would be interesting to investigate how imperfect detection affects other types of measurements, such as unambiguous or error-free easurements~\cite{rehacek}. 
If the detection is imperfect, then it may not be possible to distinguish some states unambiguously anymore. We should then instead consider a maximum confidence measurement~\cite{maxconf}. 
Also, errors in an experimental realisation will not only come from imperfections in the final detection, but necessarily also from errors in operations on the system to be measured. Finding optimal measuements for these cases will be the subject of further work.

The author gratefully acknowledges useful discussions with Mark Hillery and Mario Ziman, and financial support from EPSRC EP/G009821/1.


\begin{thebibliography}{00}
\bibitem{helstrom} C. W. Helstrom, {\it Quantum detection and estimation theory} (Academic, New York, 1976).
\bibitem{rehacek} 
J. A. Bergou, U. Herzog, and M. Hillery, Ch. 11 in
{\it Quantum state estimation} (Springer, Berlin, 2004); A. Chefles, {\it ibid.}, Ch. 12.
\bibitem{steve} S. M. Barnett, {\it Quantum information} (Oxford University Press, Oxford, 2009).

\bibitem{enk} S. J. van Enk, Pys. Rev. A {\bf 66}, 042313 (2002).

\bibitem{aspachs} M. Aspachs, G. Adesso and I. Fuentes, Phys. Rev. Lett. {\bf 105}, 151301 (2010).

\bibitem{huttner} B. Huttner {\it et al.}, 
Phys. Rev. A {\bf 54}, 3783 (1996).

\bibitem{clarkeunamb} R. B. M. Clarke {\it et al.}, 
Phys. Rev. A {\bf 63}, 040305 (2001)

\bibitem{clarkesymm} R. B. M. Clarke {\it et al.}, 
Phys. Rev. A {\bf 64}, 012303 (2001).

\bibitem{mizuno} J. Mizuno {\it et al.}, 
Phys. Rev. A {\bf 65}, 012315 (2001).

\bibitem{superadd} M. Fujiwara {\it et al.}, 
Phys. Rev. Lett. {\bf 90}, 167906 (2003).

\bibitem{source} Y. Mitsumori {\it et al.}, 
Phys. Rev. Lett. {\bf 91}, 217902 (2003). 

\bibitem{steinberg} M. Mohseni, A. M. Steinberg, and J. Bergou, Phys. Rev. Lett. {\bf 93}, 200403 (2004).

\bibitem{mosley} P. J. Mosley {\it et al.}, 
Phys. Rev. A {\bf 77}, 012113 (2008).

\bibitem{stuttgart} G. Waldherr, A. Dada, F. Jelezko, E. Andersson, and J. Wrachtrup, in preparation.

\bibitem{atompom} S. Franke-Arnold {\it et al.}, 
Phys. Rev. A {\bf 63}, 052301 (2001).

\bibitem{atompom2} E. Andersson, Phys. Rev. A {\bf 64}, 032303 (2001).

\bibitem{cavity} A. Dada {\it et al.}, 
Phys. Rev. A {\bf 83}, 042339 (2011).

\bibitem{hamilton} C. S. Hamilton {\it et al.}, 
Phys. Rev. A {\bf 79}, 023808 (2009).

\bibitem{chinese} G. Wang ang M. Ying, arXiv:0608235.
\bibitem{treepom} E. Andersson and D. K. L. Oi, Phys. Rev. A {\bf 77}, 052104 (2008).
\bibitem{fieldionization} H. Walther {\it et al.}, 
Rep. Prog. Phys. {\bf 69}, 1325 (2006).

\bibitem{mirrorsym} E. Andersson {\it et al.}, 
Phys. Rev. A {\bf 65}, 052308 (2002).

\bibitem{chou} C.-L. Chou, Phys. Rev. A {\bf 70}, 062316 (2004).

\bibitem{sych} D. Sych and G. Leuchs, New. J. Phys. {\bf 12}, 053019 (2010).

\bibitem{maxconf} S. Croke {\it et al.}, 
Phys. Rev. A {\bf 72}, 052116 (2005).
\end{thebibliography}
\end{document}